\documentclass[fleqn,usenatbib]{mnras}

\usepackage[T1]{fontenc}
\usepackage{ae,aecompl}
\usepackage{xcolor}
\usepackage{pdflscape}
\usepackage{graphicx}	% Including figure files
\usepackage{amsmath}	% Advanced maths commands
\usepackage{amssymb}	% Extra maths symbols
\usepackage{physics}
\usepackage{placeins}
\usepackage{cleveref} 
\usepackage[swedish, english]{babel}
\usepackage{newtxtext,newtxmath}
\usepackage{comment}
\newcommand{\ang}{\,\textup{\AA}}
\newcommand{\Msol}{\,M_{\odot}}
\def\feff{f_*^{\rm{eff}}}
\def\fs{f_*}
\def\fej{f_*^{\rm{ej}}}

\graphicspath{ {figures/} }

\title[Balmer breaks in high-z galaxies]{To be, or not to be: Balmer breaks in high-z galaxies with JWST}

\author[Anton Vikaeus et al.]{Anton Vikaeus$^{1}$\thanks{E-mail: anton.vikaeus@physics.uu.se},
Erik Zackrisson$^{1}$,
Stephen Wilkins$^{2,3}$,
Armin Nabizadeh$^{1}$,
Vasily Kokorev$^{4}$,
\newauthor
Abdurro'uf$^{5,6}$,
Larry D. Bradley$^{6}$,
Dan Coe$^{6,5,7}$,
Pratika Dayal$^{4}$,
Massimo Ricotti$^{8}$.
\\
$^{1}$Observational Astrophysics, Department of Physics and Astronomy, Uppsala University, Uppsala, Sweden \\
$^{2}$Astronomy Centre, University of Sussex, Falmer, Brighton BN1 9QH, UK\\
$^{3}$Institute of Space Sciences and Astronomy, University of Malta, Msida MSD 2080, Malta \\
$^{4}$ Kapteyn Astronomical Institute, University of Groningen, 9700 AV Groningen, The Netherlands\\ 
$^{5}$ Center for Astrophysical Sciences, Department of Physics and Astronomy, The Johns Hopkins University, 3400 N Charles St. Baltimore, MD 21218, USA\\
$^{6}$ Space Telescope Science Institute (STScI), 3700 San Martin Drive, Baltimore, MD 21218, USA\\ $^{7}$ Association of Universities for Research in Astronomy (AURA), Inc.~for the European Space Agency (ESA)\\ 
$^{8}$ Department of Astronomy, University of Maryland, College Park, 20742, USA\\
}
\date{Accepted XXX. Received YYY; in original form ZZZ}

\pubyear{2023}

\begin{document}
\label{firstpage}
\pagerange{\pageref{firstpage}--\pageref{lastpage}}
\maketitle

\newcommand{\LCDM}{$\Lambda$CDM}

\newcommand{\redbf}[1]{{\color{red}\bf #1 \color{black}}}

\newcommand{\ny}{$\tilde {\rm n}$}
\newcommand{\about}{$\sim$}
\newcommand{\appr}{$\approx$}
\newcommand{\gt}{$>$}
\newcommand{\um}{$\mu$m}
\newcommand{\uJy}{$\mu$Jy}
\newcommand{\sig}{$\sigma$}
\newcommand{\Lya}{Lyman-$\alpha$}
\renewcommand{\th}{$^{\rm th}$}
\newcommand{\lam}{$\lambda$}

\newcommand{\tentothe}[1]{$10^{#1}$}
\newcommand{\tentotheminus}[1]{$10^{-#1}$}
\newcommand{\e}[1]{$\times 10^{#1}$}
\newcommand{\en}[1]{$\times 10^{-#1}$}
\newcommand{\cgsfluxunits}{erg$\,$s$^{-1}\,$cm$^{-2}$}
\newcommand{\linefluxunits}{\tentotheminus{20} \cgsfluxunits}

\newcommand{\logU}{$\log(U)$}
\newcommand{\logOH}{12+log(O/H)}

\newcommand{\sinv}{s$^{-1}$}

\newcommand{\footnoteurl}[1]{\footnote{\url{#1}}}

\newcommand{\tnm}[1]{\tablenotemark{#1}}
\newcommand{\super}[1]{$^{\rm #1}$}
\newcommand{\supa}{$^{\rm a}$}
\newcommand{\supb}{$^{\rm b}$}
\newcommand{\supc}{$^{\rm c}$}
\newcommand{\supd}{$^{\rm d}$}
\newcommand{\supe}{$^{\rm e}$}
\newcommand{\supf}{$^{\rm f}$}
\newcommand{\supg}{$^{\rm g}$}
\newcommand{\suph}{$^{\rm h}$}
\newcommand{\supi}{$^{\rm i}$}
\newcommand{\supj}{$^{\rm j}$}
\newcommand{\supk}{$^{\rm k}$}
\newcommand{\supl}{$^{\rm l}$}
\newcommand{\supm}{$^{\rm m}$}
\newcommand{\supn}{$^{\rm n}$}
\newcommand{\supo}{$^{\rm o}$}

\newcommand{\squared}{$^2$}
\newcommand{\cubed}{$^3$}

\newcommand{\sqarcmin}{arcmin\squared}

\newcommand{\supcomma}{$^{\rm ,}$}

\newcommand{\rhalf}{$r_{1/2}$}

\newcommand{\chisq}{$\chi^2$}

\newcommand{\Zgas}{$Z_{\rm gas}$}  % gas-phase metallicity
\newcommand{\Zstar}{$Z_*$}  % stellar metallicity

\newcommand{\per}{$^{-1}$}
\newcommand{\inv}{\per}
\newcommand{\Mstar}{$M^*$}
\newcommand{\Lstar}{$L^*$}
\newcommand{\phistar}{$\phi^*$}

\newcommand{\logM}{log($M_*$/\Msun)}

\newcommand{\LUV}{$L_{UV}$}
\newcommand{\MUV}{$M_{UV}$}

\newcommand{\Msun}{$M_\odot$}
\newcommand{\Lsun}{$L_\odot$}
\newcommand{\Zsun}{$Z_\odot$}

\newcommand{\Mvir}{$M_{vir}$}
\newcommand{\Mt}{$M_{200}$}
\newcommand{\Mf}{$M_{500}$}

\newcommand{\Ndotion}{$\dot{N}_{\rm ion}$}
\newcommand{\xiion}{$\xi_{\rm ion}$}
\newcommand{\logxiion}{log(\xiion)}
\newcommand{\fesc}{$f_{\rm esc}$}

\newcommand{\XHI}{$X_{\rm HI}$}
\newcommand{\XHII}{$X_{\rm HII}$}
\newcommand{\RHII}{$R_{\rm HII}$}

\newcommand{\Halpha}{H$\alpha$}
\newcommand{\Hbeta}{H$\beta$}
\newcommand{\Hgamma}{H$\gamma$}
\newcommand{\Hdelta}{H$\delta$}
\newcommand{\Halphaw}{\Halpha\,$\lambda$6563}
\newcommand{\Hbetaw}{\Hbeta\,$\lambda$4861}
\newcommand{\Hgammaw}{H$\gamma$\,$\lambda$4340}
\newcommand{\Hdeltaw}{H$\delta$\,$\lambda$4101}
\newcommand{\Ha}{\Halpha}
\newcommand{\Hb}{\Hbeta}

\newcommand{\I}{\,{\sc i}}
\newcommand{\II}{\,{\sc ii}}
\newcommand{\III}{\,{\sc iii}}
\newcommand{\IV}{\,{\sc iv}}
\newcommand{\V}{\,{\sc v}}
\newcommand{\VI}{\,{\sc vi}}
\newcommand{\VII}{\,{\sc vii}}
\newcommand{\VIII}{\,{\sc viii}}

\newcommand{\HI}{H\,{\sc i}}
\newcommand{\HII}{H\,{\sc ii}}
\newcommand{\HeI}{He\,{\sc i}}
\newcommand{\HeII}{He\,{\sc ii}}

\newcommand{\CII}{[C\,{\sc ii}]}
\newcommand{\CIIw}{\CII\,$\lambda$2325 (blend)}
\newcommand{\CIII}{[C\,{\sc iii}]}
\newcommand{\CIIIw}{\CIII\,$\lambda$1908}
\newcommand{\CIIId}{C\,{\sc iii}]}
\newcommand{\CIIIdw}{C\,{\sc iii}]\,$\lambda\lambda$1907,1909}
\newcommand{\CIV}{C\,{\sc iv}}
\newcommand{\CIVw}{\CIV\,$\lambda$1549}
\newcommand{\OII}{[O\,{\sc ii}]}
\newcommand{\OIIw}{\OII\,$\lambda$3727}
\newcommand{\OIIdw}{\OII\,$\lambda\lambda$3727,3729}
\newcommand{\OIII}{[O\,{\sc iii}]}
\newcommand{\OIIIw}{\OIII\,$\lambda$5007}
\newcommand{\OIIIww}{\OIII\,$\lambda$4959,$\lambda$5007}
\newcommand{\OIIIwa}{\OIII\,$\lambda$4363}
\newcommand{\OIIIwc}{\OIII\,$\lambda$4959}
\newcommand{\NeIII}{[Ne\,{\sc iii}]}
\newcommand{\NeIIIw}{\NeIII\,$\lambda$3869}
\newcommand{\NeIIIwb}{\NeIII\,$\lambda$3968}
\newcommand{\HeIw}{HeI\,$\lambda$3889}
\newcommand{\HeIwa}{HeI\,$\lambda$4473}
\newcommand{\HeIIw}{HeII\,$\lambda$1640}
\newcommand{\NII}{[N\,{\sc ii}]}
\newcommand{\NIII}{N\,{\sc iii}]}
\newcommand{\NIV}{N\,{\sc iv}]}
\newcommand{\NIIIw}{\NIII\,$\lambda$1748}
\newcommand{\NIVw}{\NIV\,$\lambda$1486}
\newcommand{\MgII}{Mg\,{\sc ii}}
\newcommand{\MgIIw}{\MgII\,$\lambda$2800}

\newcommand{\Lyaw}{Ly$\alpha$\,$\lambda$1216}

%\NIVw, \CIVw, \HeIIw, \NIIIw, \MgIIw.

% Stark17:
%\newcommand{\ciiid}{\hbox{[C\,{\sc iii}]$\lambda1907$+C\,{\sc iii}]$\lambda1909$}}
%\newcommand{\ciiit}{\hbox{C\,{\sc iii}]$\lambda1908$}}
%\newcommand{\Hii}{\mbox{H\,{\sc ii}}}

\newcommand{\Om}{\Omega_{\rm M}}
\newcommand{\OL}{\Omega_\Lambda}

\newcommand{\etal}{et al.}

\newcommand{\citeps}{\citep}

\newcommand{\HST}{{\em HST}}
\newcommand{\SST}{{\em SST}}
\newcommand{\Hubble}{{\em Hubble}}
\newcommand{\Spitzer}{{\em Spitzer}}
\newcommand{\Chandra}{{\em Chandra}}
\newcommand{\JWST}{{\em JWST}}
\newcommand{\Planck}{{\em Planck}}

\newcommand{\Bradac}{{Brada\v{c}}}

\newcommand{\citepeg}[1]{\citep[e.g.,][]{#1}}

\newcommand{\range}[2]{\! \left[ _{#1} ^{#2} \right] \!}  % range of values in brackets

\newcommand{\grizli}{\textsc{grizli}}
\newcommand{\eazypy}{\textsc{eazypy}}
\newcommand{\msaexp}{\textsc{msaexp}}
\newcommand{\trilogy}{\textsc{trilogy}}
\newcommand{\bagpipes}{\textsc{bagpipes}}
\newcommand{\beagle}{\textsc{beagle}}
\newcommand{\photutils}{\textsc{photutils}}
\newcommand{\SEP}{\textsc{sep}}
\newcommand{\piXedfit}{\textsc{piXedfit}}
\newcommand{\pyneb}{\textsc{pyneb}}
\newcommand{\HIIC}{\textsc{hii-chi-mistry}}
\newcommand{\astropy}{\textsc{astropy}}
\newcommand{\astrodrizzle}{\textsc{astrodrizzle}}
\newcommand{\multinest}{\textsc{multinest}}
\newcommand{\cloudy}{\textsc{Cloudy}}
\newcommand{\jdaviz}{\textsc{Jdaviz}}
\newcommand\code[1]{\textsc{\MakeLowercase{#1}}}

\renewcommand{\tt}[1]{\texttt{#1}}

\newcommand{\SE}{\tt{SourceExtractor}}

\newcommand{\PD}[1]{\textcolor{blue}{[PD: #1\;]}}

\newcommand{\JD}{MACS0647$-$JD}

% Abstract of the paper
\begin{abstract}
Standard models of structure formation allow us to predict the cosmic timescales relevant for the onset of star formation and the assembly history of galaxies at high redshifts ($z>10$). The strength of the Balmer break represents a well-known diagnostic of the age and star formation history of galaxies, which enables us to compare observations with contemporary simulations -- thus shedding light on the predictive power of our current models of star formation in the early universe. Here, we measure the Balmer break strength for 23 spectroscopically confirmed galaxies at redshifts 6 $\lesssim z \lesssim 12$ using public \textit{JWST} NIRSpec data from the cycle 1 GO 1433 and GO 2282 programs (PI Coe), as well as public spectroscopic data from the JWST Deep Extragalactic Survey (JADES). We find that the range of observed Balmer break strengths agree well with that of current simulations given our measurement uncertainties. No cases of anomalously strong Balmer breaks are detected, and therefore no severe departures from the predictions of contemporary models of star formation. However, there are indications that the number of outliers in the observed distribution, both in direction of strong and weak Balmer breaks, is higher than that predicted by simulations.
\end{abstract}

% Don't make up new ones.

%%%%%%%%%%%%%%%%%%%%%%%%%%%%%%%%%%%%%%%%%%%%%%%%%%

%%%%%%%%%%%%%%%%% BODY OF PAPER %%%%%%%%%%%%%%%%%%

\section{Introduction}
 When estimating the star-forming age of an evolved galaxy, one of the more powerful proxies is the strength of the so-called Balmer break. This apparent discontinuity in the spectra of galaxies appears as a result of the complete ionization of hydrogen atoms occupying the second excited atomic state -- producing a break in the restframe continuum emission around a wavelength of $\sim 3600 \ang$. The strength of this continuum break is primarily governed by stellar physics which identifies stars of spectral type A as the dominant stellar component responsible for strong Balmer breaks in galaxy spectra. As such, the Balmer break will evolve with time due to stellar evolution and the galaxy's star-forming history. The break grows strongest at around 0.3--1\,Gyr for a simple, single-age stellar population, while leveling out as the galaxy continues to age \citep{2006Kriek}. Furthermore, in galaxies with ongoing star formation, the continuous replenishing of young stars enhances the integrated flux at shorter wavelengths, resulting in a bluer spectrum with a less pronounced Balmer break. In normal circumstances, one therefore generally requires a mature stellar population with an age of $\gtrsim 0.3$\,Gyr that dominates the integrated flux in order to expect a significant Balmer break \citep[e.g,][]{2023Steinhardt}.

The Balmer break is also sensitive to nebular reprocessing of the stellar radiation which becomes evident in cases of ongoing star formation and young stellar populations. The nebular continuum emission strongly enhances the flux at wavelengths blueward of the break -- resulting in a smaller Balmer break ratio. Furthermore, dust attenuation has the opposite effect such that a significant dust component enhances the Balmer break due to reddening \citep{2023Wilkins}.

 \begin{figure*} 
  \centering
    \includegraphics[width=\columnwidth]{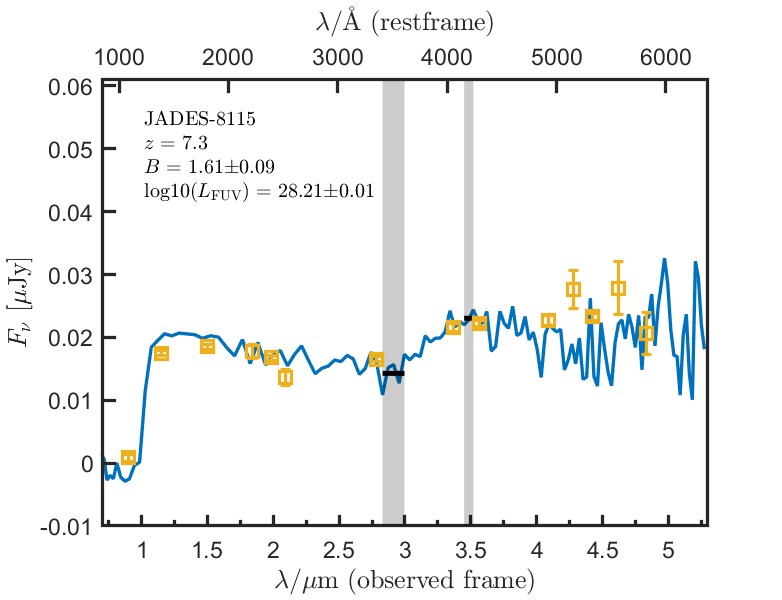}
    \includegraphics[width=\columnwidth]{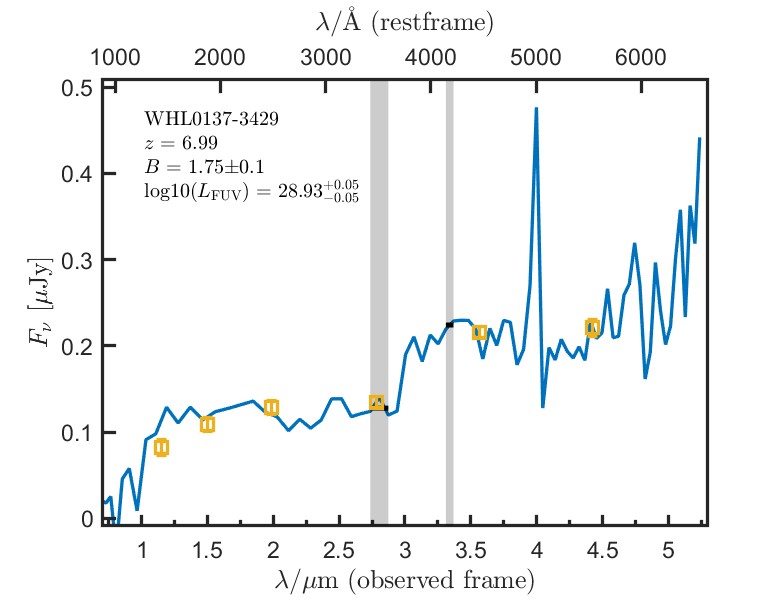}
    \caption{JWST NIRSpec/PRISM spectra and NIRCam photometry (yellow boxes) for two of the high-redshift galaxies (JADES-8115 and WHL0137-3429) from our sample with notable Balmer breaks. The gray shaded area marks the two wavelength ranges used to calculate the Balmer break. Inside the shaded regions, two solid black lines indicate the continuum level fitted and used for the Balmer break calculations.}
  \label{fig_spectras}
  \end{figure*}

The Balmer break, in conjunction with detections of strong Balmer emission lines, such as H$\alpha$ or H$\beta$, makes it possible to constrain the star formation history and the overall age of the stellar population making up a galaxy. The utility of the Balmer break as an age indicator is commonly used when analyzing spectra of galaxies at lower redshifts, where an evolved stellar population is likely a more significant component in the integrated galaxy spectra. Reaching towards the very high redshifts ($z\gtrsim 10$) now probed by \textit{The James Webb Space Telescope} (\textit{JWST}) suggests a theoretically declining strength in the Balmer break due to the predominantly young stellar populations that formed close to the onset of galaxy formation -- thus encouraging our endeavour for observational confirmation of this very assumption.

Recent discoveries \citep{2020Roberts-Borsani,2021Laporte} have complicated this picture due to the identification of $z\approx7$--9 galaxies that, if based solely on the Balmer break, argue for an onset of star formation very early in the Universe through short ($\sim 10-100$ Myr) bursts of star formation followed by quiescent phases or very low star formation rates. \citet{2018Hashimoto} provided additional evidence for such a $z\approx 9$ galaxy (MACS1149-JD1), showing a strong Balmer break that suggests the galaxy likely formed the bulk of its stars around 250 million years after the Big Bang. Whether such bursts and following quiescent phases are typical for the assembly of the first galaxies, is currently being investigated by \textit{JWST} which enables us to study their star formation histories back to the very first 100\,Myrs of cosmic time. This was just recently subject to investigation by new \textit{JWST} data, arguing for a much weaker Balmer break in MACS1149-JD1, and thus a less extreme star formation history \citep{bradač2023,2023Stiavelli}.

As of now, a few observations by \textit{JWST} consistent with Balmer breaks at high redshift have been presented. Photometrically, \citet{2023Labbe} provides evidence of a Balmer break in six galaxies with $7.4 \leq z \leq 9.1$. However, at the same time, objects with similar photometric signatures have been linked with active galactic nuclei \citep{2023LabbeAGN,2023Kokorev}, revealing the challenge of photometrically identifying clear and unambiguous Balmer breaks. So far, spectroscopic confirmation of a high-redshift galaxy with a distinct Balmer break has only been reported by \citet{2023Looser} at $z=7.3$. More recently, \citet{2023Curtis-Lake} spectroscopically confirmed multiple galaxies at extreme redshifts $z \approx 10$--13, but ruled out the existence of any prominent Balmer breaks in the data presented. \citet{2023Trussler} presents photometric data for a large number of galaxies at $7<z<12$ where they search for an excess in the F444W and F356W NIRCam filters of \textit{JWST} as a means to identify candidate high-z Balmer break galaxies. Similarly \citep{2023AtekUncover} reports three galaxies at $z\sim 9.5-10.2$ with a strong F444W excess.

A key aspect in determining the star-forming age of a galaxy from the Balmer break is the assumptions made regarding the star formation history and chemical enrichment. Recent high-redshift simulations such as FIRE (Feedback in realistic environments) by \citet{2020Ma}, FLARES (First light and reionization epoch simulations) by \citet{Vijayan2020,Lovell2020}, the SPHINX simulations \citep{2018Rosdahl,2021Katz} as well as the simulations in \citet{2023Garcia}, provide a picture different than the perhaps oversimplified models assuming a constant, or slowly varying star formation rate. Feedback processes, chemical enrichment, magnetic fields, supernova explosions, etc. all play significant roles in the way star formation proceeds after onset. Oversimplifications would evidently render our predictions less accurate and could therefore call for some revision of our contemporary models of star formation -- advocating a more detailed treatment which at this point in time is only amenable to numerical simulations such as those mentioned above.

In this work, we present spectroscopically-determined Balmer break strengths for a sample of 23 galaxies studied with \textit{JWST} and explore the agreement of these measurements with Balmer breaks derived for simulated galaxies from the First Light And Reionization Epoch Simulations \citep[FLARES;][]{Vijayan2020,Lovell2020,2023Wilkins} and DELPHI \citep{dayal2014a, dayal2022, mauerhofer2023} simulation suites. 

The paper is structured as follows: In Sect.~\ref{method}, we outline our methodology for measuring the Balmer break strength and introduce the theoretical and observational data used in this study. In Sect.~\ref{results} we present our observational findings regarding the Balmer break of high-redshift galaxies while in Sect.~\ref{discussion_conclusion} we discuss and summarize our conclusions. A flat $\Lambda$CDM cosmological model with $H_0=67.3\mathrm{\, km \, s^{-1}Mpc^{-1}}$, $\Omega_{\Lambda}=0.685$, $\mathrm{\Omega_{m}=0.315}$ and $\mathrm{\Omega_b=0.0487}$ \citep{2014Planck} is adopted throughout the paper.

\begin{table*}
	\centering
	\caption{The observed Balmer break strength $B$ measured for 23 galaxies. MACS0647 and WHL0137 IDs are from the v4 catalogue (v7 IDs in parentheses). The redshifts for the MACS0647 and WHL0137 galaxy cluster fields are fitted using \msaexp, while the spectra and estimated redshifts for the galaxies in the JADES survey are retrieved from \citet{2023Bunker}. The magnification from gravitational lensing $\mu$ is estimated from the lens models discussed in sec.~\ref{subsec:Observations}. The total stellar mass ($M_\star$) is fitted using piXedfit \citep{2021Abdurrouf}, while the far ultraviolet luminosity $(L_\mathrm{FUV})$ is calculated from the spectra using the restframe continuum at $1500 \ang$. The NIRCam F200W filter flux is uncorrected for lensing.}
	\begin{tabular}{l c c c c c c c c} % five columns, alignment for each
		\hline \hline 
		ID & RA & Dec &  $B= \frac{F_\nu (4200\ang)}{F_\nu (3500\ang)}$ & $z_\mathrm{spec}$ & $\mu$ & log$_ {10} M_\star$ & log$_ {10}L_\mathrm{FUV}$ & F200W \\
     \textit{v4} \quad \textit{(v7)} &deg  &deg & & & & ($\Msol$) & (erg/s/Hz)  & nJy \\
		\hline
    \textbf{MACS0647} &  &  &  &   &  &  &  & \\
		% 3349 & $0.65 \pm 0.03$ & 10.17 & -- & 29.93 (mag?)  \\
		 3568 (4922) &101.903980 &70.242978 & $0.87 \pm 0.25$ & 9.25 & $2.5^{+0.1}_{-0.1}$  & $8.991^{+0.659} _{-0.649}$  & $28.30^{+0.03} _{-0.03}$ & $28.5\pm 2.2$\\
		 3754 (5191) &101.920529  &70.244899 &$0.73 \pm 0.13$ & 7.47 & $6.3^{+0.5}_{-0.4}$ & $8.771^{+0.544} _{-0.493}$  & $28.45^{+0.04} _{-0.04}$ & $216\pm 5.1$ \\
		 1715 (2121) &101.983171 &70.209787&$1.05 \pm 0.15$ & 6.14& $1.1^{+0.1}_{-0.1}$  & $9.370^{+0.588} _{-0.617}$ & $28.23^{+0.05} _{-0.05} $  & $37.6\pm 4.0$\\
      		3989 (5591)  &101.940754  &70.249101 & $1.12 \pm 0.15$ & 6.14& $1.7^{+0.1}_{-0.1}$ & $9.398^{+0.577} _{-0.670}$ & $27.51^{+0.04} _{-0.04} $ & $7.6\pm 2.6$ \\
   		3308 (4533) &101.952449 &70.239294 & $1.13 \pm 0.05$ & 6.13& $2.5^{+0.2}_{-0.1}$ & $8.838^{+0.597} _{-0.497}$ & $27.94^{+0.03} _{-0.04}$ & $68.4\pm 3.0$\\
		3208 (4411)  &101.918791 &70.237614 & $1.41 \pm 0.13$ & 6.12& $1.9^{+0.1}_{-0.1}$ & $9.372^{+0.606} _{-0.470}$ & $28.59^{+0.04} _{-0.04}$ & $144\pm 4.2$\\
		\hline
  \textbf{WHL0137} &  &  &  &   & &  &    \\
  	 1968 (5609) &24.355077 &-8.447929 & $0.96 \pm 0.18$ & 8.22 & $8.6^{+14}_{-6}$  & $7.969^{+1.095} _{-0.921}$  & $27.92^{+0.57} _{-0.44}$ & $72.3\pm 7.4$\\
   %   1322 & & & $1.03 \pm 0.25$ & 7.27  & $7.372^{+0.561} _{-0.594}$  & $28.89^{+0.10} _{-0.20}$\\
  	 3429 (7424) &24.324384 &-8.418989 & $1.75 \pm 0.10$ & 6.99 & $1.1^{+0.1}_{-0.1}$ & $9.875^{+0.573} _{-0.522}$  & $28.93^{+0.50} _{-0.50}$ & $128.4\pm 7.8$\\
		\hline
  \textbf{JADES} &  & & & & & &    \\
 	 10014220 &53.16477 &-27.77463& $1.25 \pm 0.44$ & 11.58 & -- & -- & $28.39 \pm 0.03$ & $14.6 \pm 0.7$ \\
		 10014177 &53.15884 &-27.77349 & $0.86 \pm 0.34$ & 10.38& --  & -- & $28.09 \pm 0.02$ & $7.7\pm 0.6$ \\
		 6438  &53.16735 &-27.80750 & $0.67 \pm 0.15$ & 9.70& -- & -- & $28.36 \pm 0.01$ & $19.1\pm 0.6$\\
    10058975  &53.11243  &-27.77461 &$0.74 \pm 0.06$ & 9.43& -- & -- & $28.80 \pm 0.01 $ & -- \\
    8013  &53.16446 &-27.80218  &$0.65 \pm 0.35$ & 8.47& -- & -- & $28.03 \pm 0.04$ & $12.1\pm 0.8$\\
    21842  &53.15683  &-27.76716 & $0.73 \pm 0.13$ & 7.98 & --& -- & $28.13 \pm 0.01 $ & $11.5\pm 0.6$\\
    8115  &53.15508 &-27.80177 &$1.61 \pm 0.09$ & 7.30& -- & -- & $28.21 \pm 0.01$ & $16.7\pm 0.5$\\
    8079  &53.15283 &-27.80194 &$1.28 \pm 0.29$ & 7.26& -- & -- & $27.85 \pm 0.02 $ & $8.6\pm 0.5$\\
    10013905 &53.11833 &-27.76901 &  $0.71 \pm 0.16$ & 7.20 & --& -- & $28.17 \pm 0.01 $ & --\\
    4297  &53.15579 &-27.81520 & $0.80 \pm 0.22$ & 6.71& -- & -- & $28.03 \pm 0.02 $ & $10.8\pm 0.6$\\
    3334  &53.15138  &-27.81916 & $1.30 \pm 0.23$ & 6.71& -- & -- & $27.82 \pm 0.01 $ & $6.1\pm 0.8$\\
    16625  &53.16904 &-27.77884 & $0.80 \pm 0.07$ & 6.63& -- & -- & $28.10 \pm 0.01 $ & $13\pm 0.5$\\
    18179  &53.17582 &-27.77446 &$1.41 \pm 0.14$ & 6.34& -- & -- & $28.26 \pm 0.02$ & $24\pm 0.7$\\
    18846  &53.13492 &-27.77271&$0.90 \pm 0.03$ & 6.33& -- & -- & $28.59 \pm 0.01 $ & $43.5\pm 0.6$\\
    18976  &53.16660& -27.77240& $1.07 \pm 0.12$ & 6.33 & --& -- & $28.05 \pm 0.01 $ & $11.6\pm 0.6$\\
		\hline \hline
	\end{tabular}
    \label{table_BBR}
\end{table*}

\section{Methods}\label{method}
We define the strength of the Balmer break following the approach explained in \citet{Binggeli2019}, where the continuum (in units of $F_\nu$) is fitted both at wavelengths longward (at $4200 \ang$) and shortward (at $3500 \ang$) of the break. The ratio of the two continuum levels forms an index quantifying the strength of the Balmer break, i.e. $B = F_\nu (4200$\ang$) /F_\nu (3500$\ang$)$.

Entangled within the Balmer break region is also another feature known as the $4000\ang$ break. This feature has a similar effect on the spectrum and is therefore usually merged with the Balmer break into one single break. However, the mechanisms behind this break are different, with a notably stronger dependence on metallicity. The closely spaced absorption features from ionized metals reduce the flux at redder wavelengths, therefore strengthening the break feature \citep{1983Bruzual,2011Kriek}. For the purposes of high redshift observations with telescopes like \textit{JWST}, the dim nature of distant and faint objects generally does not provide sufficient resolution to resolve the two break features. Therefore, in this paper, we do not distinguish the two and instead quantify the Balmer/4000$\ang$ break through wavelength ranges covering either sides of the break region as a whole \citep{Binggeli2019,2023Wilkins}.

Utilizing the above-mentioned definition, we determine the Balmer break from spectroscopic continuum measurements, coupled to spectroscopic redshifts, obtained from \textit{JWST}/NIRSpec. This ensures that we deal with well-established redshift values for the galaxies under consideration, which allows us to constrain the Balmer break while minimizing potential contamination from strong emission lines.

 The necessity for spectroscopic studies of the Balmer break becomes evident due to the prevalent presence of strong emission lines such as [O III]$_{\lambda4959,5007}$ and H$\beta$, which can complicate the analysis of photometric data and mimic the presence of a strong Balmer break as they enter into the filter depending on redshift \citep[e.g.,][]{2023Stefanon}.

\begin{figure*} 
  \centering
    \includegraphics[width=\columnwidth]{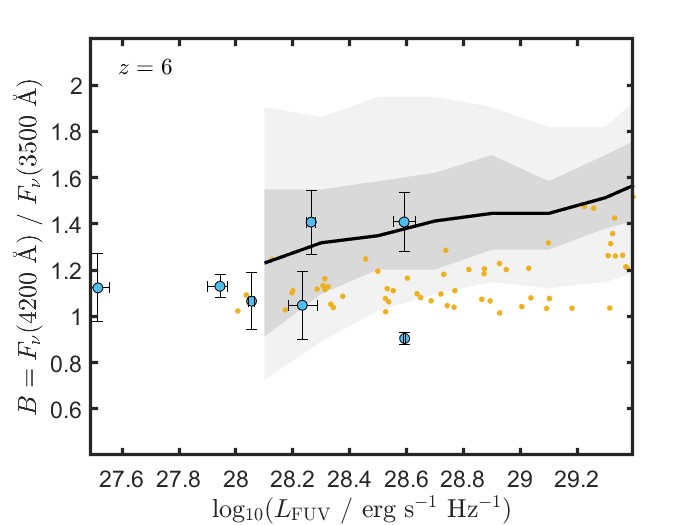}
    \includegraphics[width=\columnwidth]{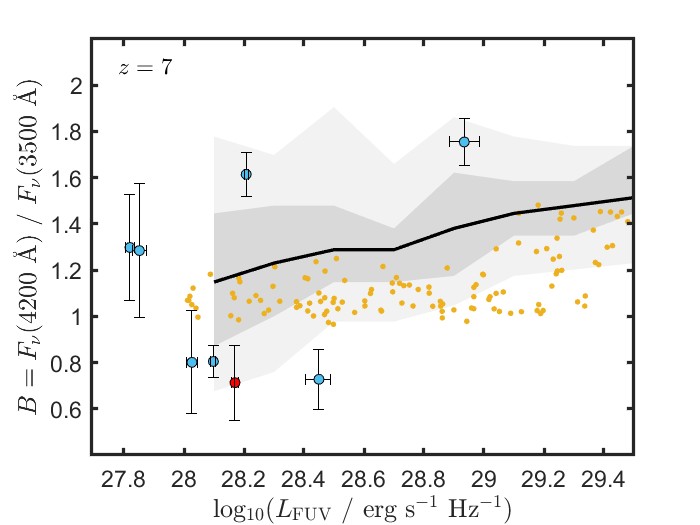}
    \caption{The Balmer break strength $B = F_\nu (4200$\ang$) /F_\nu (3500$\ang$)$ as a function of far ultraviolet luminosity $(L_\mathrm{FUV})$ at redshifts $z=6$ and $z=7$. The light blue dots with accompanying error bars represent the measured Balmer break strength derived from the observations that have spectroscopic redshifts within $z = 6 \pm 0.5$ (left) and $z = 7 \pm 0.5$ (right). Red markers indicate galaxies with either missing NIRCam photometry, or photometric data that cannot be trivially reconciled with spectroscopy. The large errors in $L_\mathrm{FUV}$ for some of the galaxies are due to gravitational lensing. The black solid line is the median Balmer break strength from the FLARES simulations, while the shaded regions correspond to the 2.2-97.8th (lighter shade) and 15.8-84.2th (darker shade) percentiles of the distribution. Orange dots correspond to simulated Balmer break strengths from the DELPHI models.}\label{fig_BBRvsLfuv_z6_z7}
  \end{figure*}
  
\subsection{Theoretical predictions of the Balmer break from simulations}
\subsubsection{FLARES}
We present the simulated Balmer break strengths derived from the FLARES simulations within the redshift range of $z=6$--10. \citet{2023Wilkins} provides a thorough analysis of the physical mechanisms affecting the Balmer break in high-z galaxies. They showed the various factors contributing to the observed strength of the break, including aspects such as star formation history (SFH), metallicity, dust, escape of Lyman-continuum (LyC) radiation, stellar initial mass function (IMF), and also the employed stellar population synthesis (SPS) models. These simulations demonstrate that a galaxy with a given far ultraviolet luminosity (or total stellar mass) may exhibit Balmer breaks that may deviate by $\sim 20$--30 percent from the median value. Considering the variations seen in the simulations, we expect that observations should frequently fall within the 2.2--97.8th percentile if our contemporary models of galaxy evolution are accurate.
\subsubsection{DELPHI}
We also compare our observational results to those from the \code{delphi} semi-analytical model \citep{dayal2014a, dayal2022, mauerhofer2023}. This model includes all the key processes of mergers and accretion in assembling the dark matter halo mass and gas mass up to $z \sim 40$, starting at $z \sim 4.5$ with a time resolution of 30 Myr and a halo mass resolution of $10^8 \Msol$. The available gas mass in any halo can form stars with an ``effective efficiency" of $\feff$, which is the minimum between the efficiency that produces enough Type II Supernova (SNII) energy to eject the remainder of the gas ($\fej$) and an upper maximum (mass- and redshift- independent) threshold ($\fs$). The models include the key processes of production, astration, destruction (of dust into metals), ejection and dust grain growth in the ISM (that leads to a corresponding decrease in the metal mass) to calculate the total dust and metal masses for each galaxy. Crucially, this model contains only two mass- and redshift-independent free parameters to match observations.  These are the maximum (instantaneous) star formation efficiency of $f_* = 8\%$ and the fraction $f_w (\sim 7.5$\%) of the SNII explosion energy that is available to drive an outflow. These parameters have been tuned to simultaneously reproduce the observed stellar mass function and the UV luminosity function at $z \sim 5-12$. The integrated spectrum for each galaxy is obtained by summing the spectrum from each burst of star formation, accounting for its metallicity, and using a Salpeter IMF between $0.1-100 \Msol$ in the \code{starburst99} \citep{leitherer1999} stellar population synthesis model. For the \code{delphi} model we consider the fiducial case of $f_\mathrm{esc}=0$ resulting in a maximal contribution from both continuum nebular emission and nebular emission lines. For nebular emission lines we use the metallicity-dependent results tabulated in \citet{anders2003}.

  \begin{figure} 
    \includegraphics[width=\columnwidth]{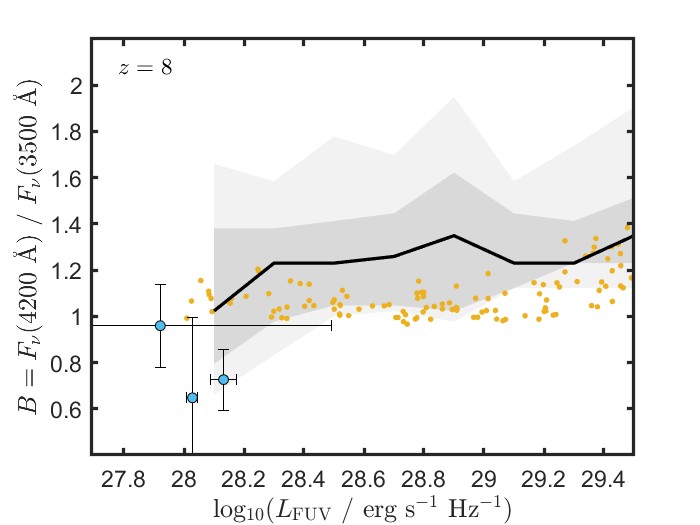}
    \caption{Same as in Fig.~\ref{fig_BBRvsLfuv_z6_z7}, here for $z=8$. Strong gravitational lensing ($\mu\sim 7.9^{+12}_{-6}$) is responsible for the wide error bars in $L_\mathrm{FUV}$ for one of the objects (identified as WHL0137-1968).}\label{fig_BBRvsLfuv_z8}
  \end{figure}
  
  \begin{figure*} 
  \centering
    \includegraphics[width=\columnwidth]{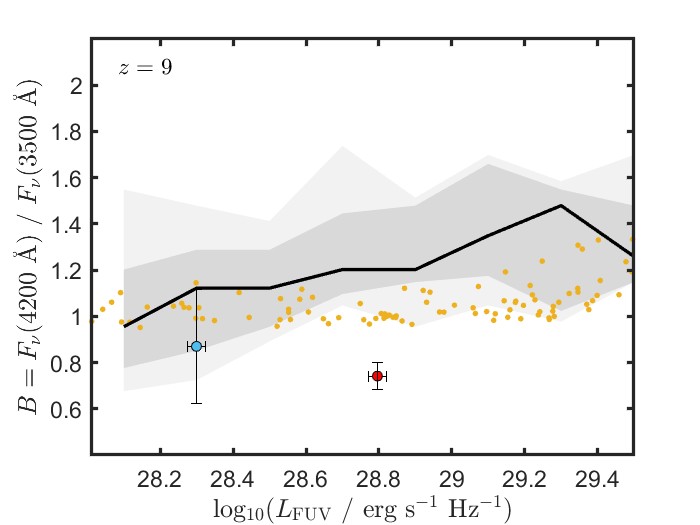}
    \includegraphics[width=\columnwidth]{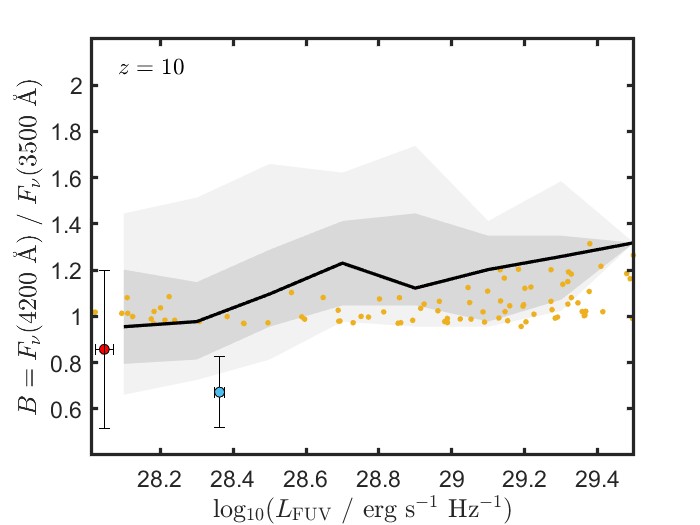}
    \caption{Same as in Fig.~\ref{fig_BBRvsLfuv_z6_z7}, here for $z=9$ and $z=10$. Spectra for the outliers at $z=9$ can be seen in fig.~\ref{fig_z9_outliers}, where the clear outlier is identified as JADES-10058975. Similarly, spectra for the $z=10$ galaxies can be seen in fig.~\ref{fig_z10_outliers}, where the weaker break is identified with JADES-6438. The red markers indicate galaxies with either missing NIRCam photometry, or, photometric data that cannot be trivially reconciled with spectroscopy.} \label{fig_BBRvsLfuv_z9_z10}
  \end{figure*}

 \begin{figure} 
    \includegraphics[width=\columnwidth]{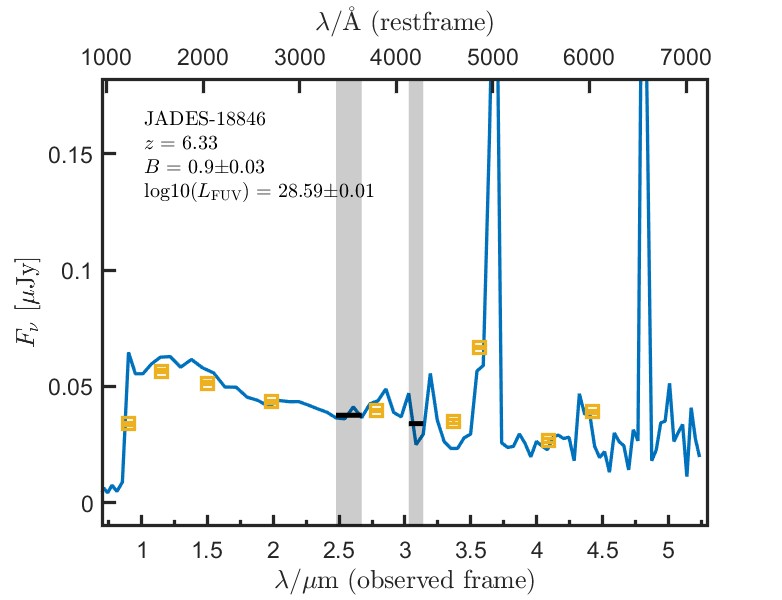}
    \caption{JWST NIRSpec/PRISM spectra and NIRCam photometry for the outlier at  $z = 6.33$, seen in the left panel of fig.~\ref{fig_BBRvsLfuv_z6_z7}. This galaxy has a quite small Balmer break strength with low error margin, placing it outside the 2.2-97.8th percentile even if considering the error. The spectrum shows that the break is extracted in a region where a notable fluctuation in the continuum shape is present. This has an impact on the calculated break strength and could therefore possibly explain the low value.} \label{fig_18846}
  \end{figure}

   \begin{figure*} 
  \centering
    \includegraphics[width=\columnwidth]{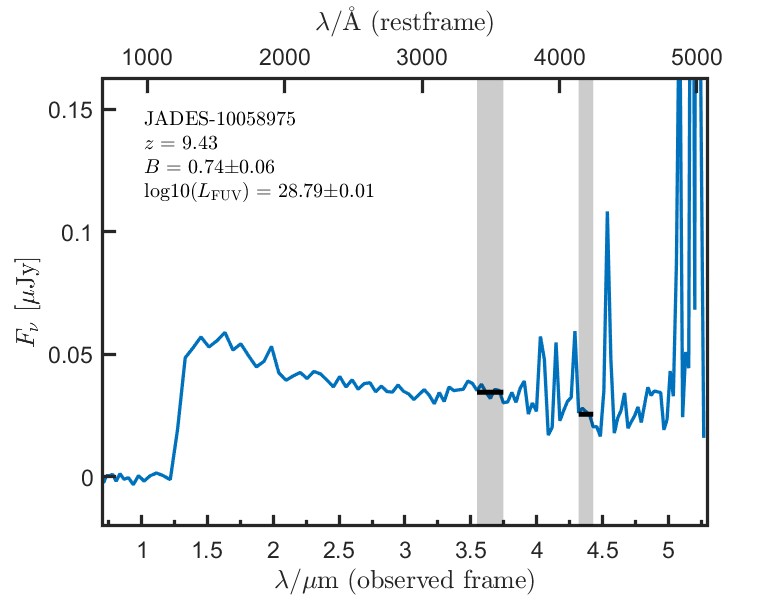}
    \includegraphics[width=\columnwidth]{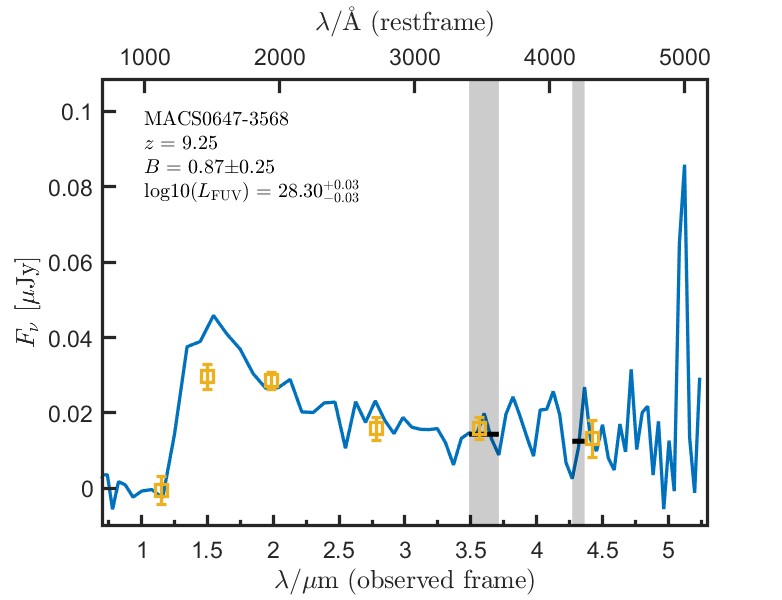}
    \caption{JWST NIRSpec/PRISM spectra and NIRCam photometry for the two $z\approx 9$ outliers in the left panel of fig.~\ref{fig_BBRvsLfuv_z9_z10}. The JADES-10058975 galaxy lacks photometric data, which raises some concern due to its significantly weak Balmer break strength ratio with a suspiciously small error margin. MACS0647-3568 lies outside of the 2.2-97.8th percentile of the FLARES simulations but has error margins that places it within the distribution and shows an adequate agreement with photometry/spectroscopy in the break region.}\label{fig_z9_outliers}
  \end{figure*}

   \begin{figure*} 
  \centering
    \includegraphics[width=\columnwidth]{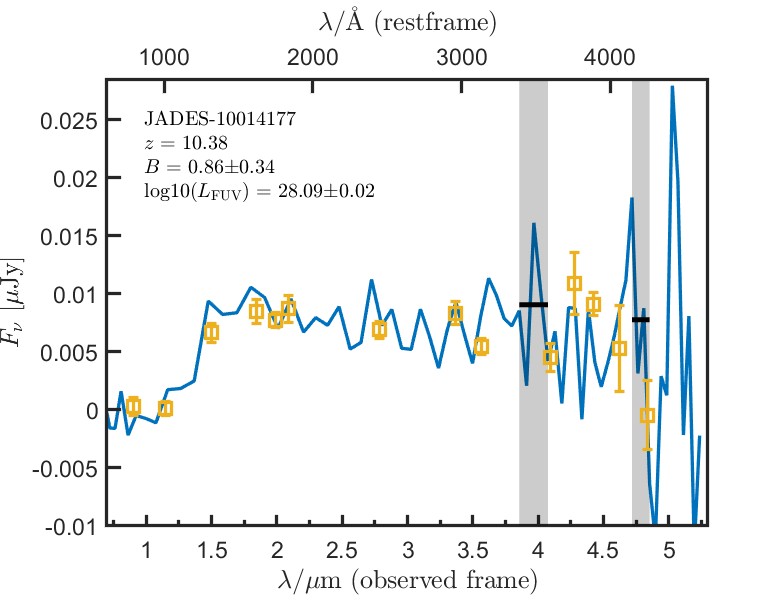}
    \includegraphics[width=\columnwidth]{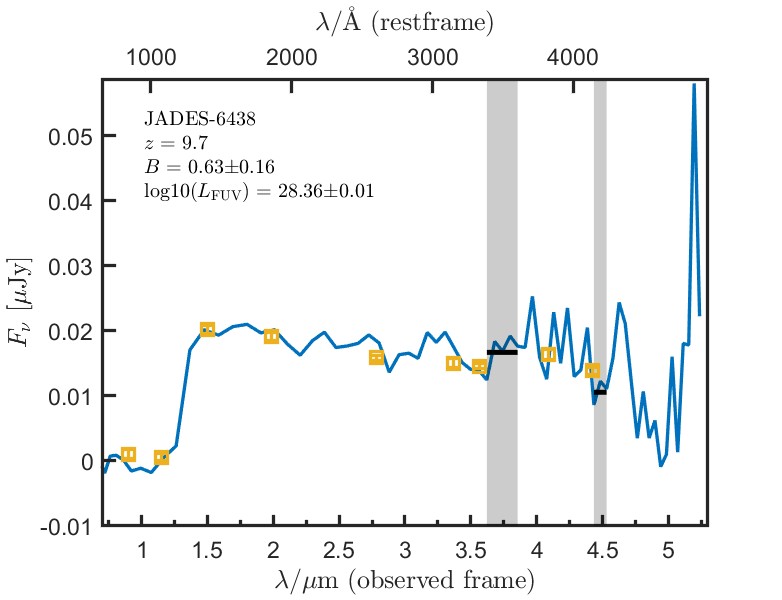}
    \caption{JWST NIRSpec/PRISM spectra and NIRCam photometry for the $z\ = 10.38$ (left) and $z\ = 9.7$ (right) galaxies seen in fig.~\ref{fig_BBRvsLfuv_z9_z10}. JADES-10014177 shows a spectrum with significant noise/variation in the Balmer break region where spectra and photometry are hard to reconcile.}\label{fig_z10_outliers}
  \end{figure*}

\subsection{NIRSpec/NIRCam observations of high-z candidates} \label{subsec:Observations}

In order to spectroscopically constrain the Balmer break in high redshift galaxies, we utilized prism data from the NIRSpec instrument on \textit{JWST}, which covers the the near-infrared wavelength range (0.6--5.3$\mu$m) and allows continuum measurements beyond the rest-wavelength Balmer break for redshifts up to $z \lesssim 11.5$.

The data analyzed here partly comes from the JWST cycle-1 GO 1433 program (PI Coe), which observed the cluster MACS J0647.7+7015 (hereafter MACS0647), and GO 2282 (PI Coe), which observed the cluster WHL0137-08 (hereafter WHL0137). These data were retrieved from MAST, which goes through STScI JWST pipeline\footnote{\url{https://github.com/spacetelescope/jwst}} version 1.9.2. Spectral lines and features are fitted using \msaexp \footnote{\url{https://github.com/gbrammer/msaexp}} version 0.6.0 to determine the redshift. Photometric data is processed through \grizli\  \citep{Grizli}\footnote{Repositories at: \url{https://dawn-cph.github.io/dja}}. 

MACS0647 and WHL0137 are known cluster lenses which implies that gravitational lensing of the observed flux is to be expected for several of the galaxies. In this dataset, we find two cases of strong gravitational lensing. The galaxy WHL0137-1968 is gravitationally lensed into an arc and estimated to have a magnification of $\mu \sim 7.9 ^{+12} _{-6}$ \citep{2022Bradley} based on the photometric redshift estimate of $z_\mathrm{ph} = 9$. With a spectroscopic redshift of $z=8.22$ we find a similar magnification estimate of $\mu \sim 8.6 ^{+14} _{-6}$ using the same lensing models as in \citet{2022Bradley}. These models include Lenstool \citep{2009Jullo}, WSLAP \citep{2005Diego,2007Diego}, glafic \citep{2010Oguri}, and Light-traces-mass \citep{2005Broadhurst,2009Zitrin,2015Zitrin}. The galaxy WHL0137-3249 is not located near the cluster core which places it in region with very minor magnifications, here we use conservative estimates of $\mu \sim 1.1 \pm 0.1$. The galaxy MACS0647-3754 is also strongly lensed as indicated in imaging. Here, we estimate a magnification $\mu \sim 6.3 ^{+0.5} _{-0.4}$ using the glafic lens model. The remaining galaxies from the MACS0647 field were estimated to have smaller magnifications, found to be in the range $1.1 \lesssim \mu \lesssim 2.5$ with errors in magnification of $\sim \pm 0.1$.  This introduces wider error bars in the calculated restframe far ultraviolet luminosity ($L_\mathrm{FUV}$), but will not be large enough (for the weaker lensing estimates) to impact our conclusions regarding the agreement between observations and the simulated distributions, shown in figures \ref{fig_BBRvsLfuv_z6_z7}-\ref{fig_BBRvsLfuv_z9_z10}. 

Additional galaxies from the JADES program are retrieved from fully reduced public data \citep{2023Bunker} with spectroscopic redshift estimates. 

In order to reduce the impact of noise and unwanted features the spectrum is rebinned, resulting in a smoother continuum from which we can calculate the Balmer break strength. This slightly affects the width (i.e., wavelength range) of the regions we picked for the calculation. However, care has been taken to make sure that parts of the spectrum that the feature strong emission lines have been left out.    

The far ultraviolet luminosity is calculated from the spectra using the restframe continuum at $1500 \ang$ after scaling the observed spectra to the photometric measurements. Scaling the spectrum also enables us to check the consistency between photometry and spectroscopy in the region where the Balmer break is measured. Discrepancies between the two would make the inferred Balmer break strength less robust, and a few such cases have been noted, which are further discussed in
section \ref{discussion_conclusion}. 

The total stellar masses for the objects in MACS0647 and WHL0137 are estimated using piXedfit \citep{2021Abdurrouf,2023Abdurrouf} and corrected for gravitational lensing. Note that we include no corresponding plots for the Balmer break strength as a function of total stellar mass in this paper \citep[see][for simulations]{2023Wilkins}. We furthermore include no total stellar mass estimates for the galaxies taken from the JADES program.

This is not an exhaustive list of all spectroscopically confirmed galaxies with possible Balmer breaks, but this data set should serve as a quantitative sample for the purposes of seeing whether present simulations are in line with observations and to rule out frequent departures from model predictions.  

Presently there are several galaxies in the literature showing significantly strong Balmer breaks. \citet{2018Hashimoto} estimates a Balmer break strength $\sim 2.3 \pm 0.4$ \citep[recently challenged by][]{bradač2023} using only slightly different wavelength ranges to define the break. \citet{2023Wilkins} also presents a Balmer break strength estimate from \citep{2023Carnall} which suggests a break strength of $\sim 2.5$, although in a galaxy at lower redshift ($z \approx 4.7$). The galaxy studied by \citet{2023Looser} is also represented in our sample (JADES-8115) where we find a clear Balmer break with a strength of $\sim 1.6$ (see fig.~ \ref{fig_spectras}). 

\begin{figure} 
    \includegraphics[width=\columnwidth]{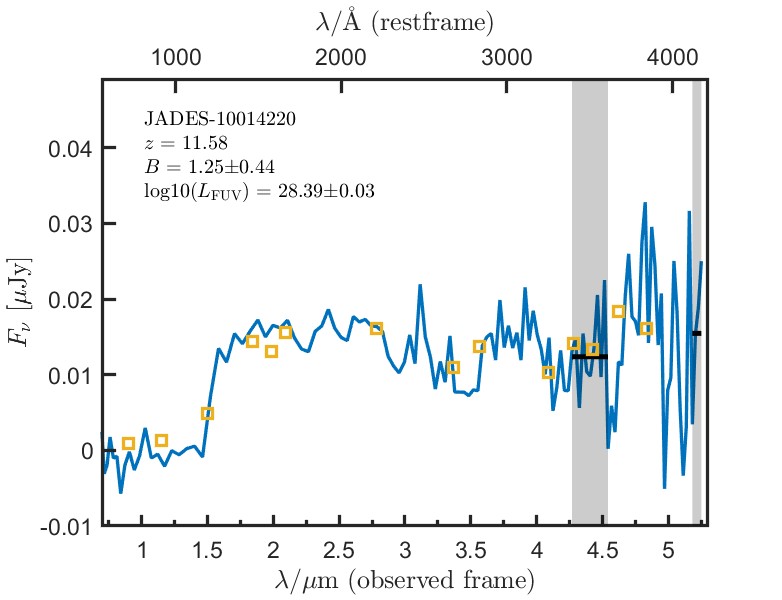}
    \caption{JWST NIRSpec/PRISM spectra and NIRCam photometry for the $z = 11.58$ galaxy in table \ref{table_BBR}. The high redshift of this galaxy makes the extraction of the Balmer break strength challenging due to severe noise at the wavelengths where the break is measured.} \label{fig_z_11_58}
  \end{figure}

We moreover include one galaxy with $z = 11.58$ (see fig.~\ref{fig_z_11_58}) for which the rest-frame 4200 {\AA} continuum data point longward of the Balmer break falls at the very edge of the NIRSpec range. The shifting of the Balmer break to these very red wavelengths combined with the low brightness introduces severe noise, which makes the extraction of reliable Balmer break measurements very challenging. No comparison to simulations is presented for this object, since we currently lack simulated data at $z\gtrsim 11$.

\section{Results}\label{results}
We assemble spectra from 23 galaxies with spectroscopically confirmed redshifts and calculated their Balmer break strength (see table \ref{table_BBR}). Based on the simulations included in this paper we find that the majority (18 out of 23) of our observed Balmer breaks can be accounted for considering the overall distribution of simulated galaxies as the calculated values fall within the shaded areas surrounding the median in figures~\ref{fig_BBRvsLfuv_z6_z7}-\ref{fig_BBRvsLfuv_z9_z10}. Considering the uncertainties in the calculated Balmer break strength we find that 19 out of 23 galaxies are consistent with the simulations. Some of the galaxies have far ultraviolet luminosities below the resolution limit of the simulations but can be extrapolated to fall within the predictions. However, we find three clear candidates (JADES-18846 and 10058975, MACS0647-3754) which show Balmer breaks that deviate significantly from the simulated data such that they are consistently lower than those predicted by the simulations, even when considering the 2.2-97.8th percentile variations from the median in FLARES and the measurement uncertainties of the break strength. Since MACS0647-3754 is gravitationally lensed with resulting large errors in $L_\mathrm{FUV}$, this outlier can potentially be alleviated, such that it falls into the distribution if sufficiently lensed. Using one lens model (glafic), we estimate that MACS0647-3754 is lensed by $\mu \sim 6.3 \pm 0.5$, which is just slightly too low to place it within the simulated distribution. The discrepancy between observation and simulation is more considerable for the higher redshifts ($z\gtrsim 8$) where we albeit have fewer observed galaxies with a well-constrained Balmer break. 

Our analysis shows that the galaxies included in the JADES program reveal an acceptable level of statistical agreement between the observed and simulated datasets. The majority of our JADES observations exhibit consistency with the simulated results for redshifts $z\gtrsim 6$. However, an exception arises in the case of JADES-10058975 (see left panel fig.~\ref{fig_z9_outliers} for spectra), showing a weak Balmer break strength of $0.74\pm0.06$ at $z\sim 9$. Considering the relatively minor uncertainty associated with this measurement, it deviates significantly from the corresponding simulations by more than 3$\sigma$ from the 97.8\% significance interval.

The galaxies in the MACS0647 and WHL0137 clusters also show good consistency with the simulations.

There are no objects in our sample with extreme Balmer break strengths $(\gtrsim 2)$ that suggest any major tensions with standard models of galaxy formation.

 \begin{figure} 
    \includegraphics[width=\columnwidth]{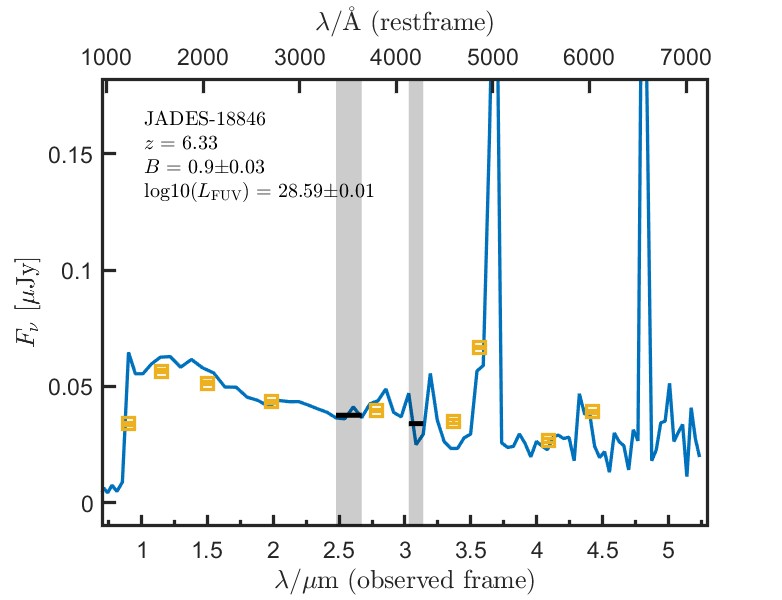}
    \caption{JWST NIRSpec/PRISM spectra and NIRCam photometry for the $z = 6.33$ galaxy in table \ref{table_BBR}. The weak Balmer break strength in comparison with simulations and small errors makes this galaxy an outlier. The spectrum reveals a significant spectral feature exactly where the break strength is measured.} \label{fig_spec_18846}
  \end{figure}

\section{Discussion and conclusion}\label{discussion_conclusion}
In this paper, we evaluated the magnitude of the Balmer break in 23 spectroscopically confirmed galaxies spanning redshifts from 6.1 to 11.6. For this, we utilized JWST/NISRSpec data obtained from the GO 1433 and GO 2282 programs together with the publicly available JADES observations. Despite our analysis revealing a reasonable agreement between the observed strength of the Balmer break and contemporary simulations, given the uncertainties inherent in our measurements, some outliers were identified when comparing to the included simulations (see~Sec.~\ref{results}). These Balmer breaks fall below the predictions by about \textbf{$60 \%$} from the median with respect to the FLARE simulations while slightly less (by $\sim$ 10 \%) than that when comparing to the median of the DELPHI models. 

Currently, other observations of galaxies at high redshifts suggest a significantly more bursty star formation history than normally expected in standard models of star formation \citep{2023Sun_a,2023Endsley_a,2023LooserBursty,2023Sun_b,2023Endsley}. Such bursty star formation implies a sporadically replenished young stellar component in the galaxy, resulting in a highly varying Balmer break strength which cannot grow particularly strong if burst reoccurs on timescales shorter than $\sim 300$ Myr. Overall, this would push the simulated Balmer breaks down and increase the scatter in figures~ \ref{fig_BBRvsLfuv_z6_z7}-\ref{fig_BBRvsLfuv_z9_z10}, and in doing so somewhat improving the agreement between models and observations.

We find no cases of extreme Balmer breaks, such as those previously found in, e.g., \citet{2018Hashimoto}. Taking our observations at face value suggests that such extreme Balmer breaks are indeed rare. However, one must acknowledge that the 23 galaxies presented in this paper do not provide a sufficiently large sample in order to draw statistically relevant conclusions regarding the observed and simulated distributions for all included redshifts. We therefore risk missing some of the important details of the true overall distribution of Balmer breaks, especially for high-redshift galaxies ($z\gtrsim 10$) where the inherent challenge of attaining high-quality spectra of significant numbers of galaxies is evident.

We conclude that FLARES \citep{2023Wilkins} make predictions that agree fairly well with observations as $\sim$ 82\% of the galaxies included in this paper agree with the predictions given that we allow reasonable uncertainties and variations from the simulated median. We do however find several accounts of particularly low Balmer break strengths, falling below the simulated median with as much as $\sim$ 60\% in the more extreme cases. Looking at the spectra of one of these galaxies (JADES-10058975, left panel fig.~\ref{fig_z9_outliers}) reveals a very blue slope and spectral features/emission lines in the break region, resulting in the weak break strength. The underlying reason for this is speculative but could be explained by a significantly young stellar population with strong nebular emission from highly ionizing stars. 

While we find reasonably good agreement between simulations and our observed galaxy population, we observe several galaxies with Balmer breaks quite far from the median predictions of FLARES. If this is the actual case, one could argue that his indicates there is some missing ingredient from the simulations related to feedback of e.g. early growth of black holes, stochastic star formation histories etc..

The predictions on the Balmer break strength from the DELPHI model have a slightly lower median and a smaller distribution than those seen in FLARES. The discrepancy between model and observation for the outliers with weaker Balmer breaks is therefore lower here. However, the objects with particularly weak Balmer break strengths are still to be considered as outliers in the distribution. Considering the lower median and more narrow distribution of the DELPHI models, we find that the galaxies with high observed Balmer break strengths (see e.g., $z=7$ in fig.~\ref{fig_BBRvsLfuv_z6_z7}) are harder to reconcile with the models.

Several galaxies show spectral features and noise that make the measurement of the Balmer break strength challenging. This can be seen clearly in, e.g., JADES-10014177 (left panel, fig.\ref{fig_z10_outliers}). The spectrum shows a lot of variation which is reflected in the relatively large error in the measured Balmer break strength. This also reveals an example where photometry and spectroscopy are hard to reconcile in order to help constrain the Balmer break. JADES-18846 (fig.~\ref{fig_spec_18846}) also reveals a scenario where a spectral feature complicates the break measurement. The estimated errors are very small due to a relatively clean spectrum in the break region, but the spectral feature pushes the break strength down slightly. The true Balmer break strength could potentially be closer to $\sim 1$, which would bring this galaxy right to the very edge of the 97.8th percentile of the simulated FLARES distribution.

While one of the aims of this paper was to utilize spectroscopy as a powerful tool to alleviate some of the obstacles related to photometric measurements of the Balmer break strength, we see that even with spectroscopic data our analysis is rarely straightforward. Due to noise and spectral features in the observed wavelength ranges corresponding to the Balmer break for high-redshift galaxies, we find few clear examples of a Balmer break (such as in fig.~\ref{fig_spectras}), but often rather measure the slope of the continuum. This, on the other hand, is still revealing of the underlying physical processes of star formation that is at play and therefore serves an important purpose when measured.

Larger observational datasets with more calculated Balmer breaks will improve the robustness of our findings. We conclude however that given our current dataset, we can find no significant deviations from predictions based on standard models of structure and star formation.

So the conspicuous title: To be or not to be; Balmer breaks in high-z galaxies with \textit{JWST}, cannot be unambiguously answered by this paper alone. While our findings indicate that the simulated predictions agree fairly well with observations we find no cases similar to the extreme Balmer breaks presented in the literature -- suggesting such cases are indeed rare.

\section*{Acknowledgements}
AV and EZ acknowledge funding from the Swedish National Space Agency. AN and EZ acknowledge funding from Olle Engkvists Stiftelse. EZ also acknowledges grant 2022-03804 from the Swedish Research Council. PD acknowledges support from the Dutch Research Council (NWO) through the award of the VIDI Grant 016.VIDI.189.162 (``ODIN") and the European Commission's and University of Groningen's CO-FUND Rosalind Franklin program.
 	
\section*{Data availability}
The data underlying this article will be shared on reasonable request to the corresponding author.
%%%%%%%%%%%%%%%%%%%%%%%%%%%%%%%%%%%%%%%%%%%%%%%%%%

%%%%%%%%%%%%%%%%%%%% REFERENCES %%%%%%%%%%%%%%%%%%

\bibliographystyle{mnras}
\bibliography{bibliography} %

%%%%%%%%%%%%%%%%%%%%%%%%%%%%%%%%%%%%%%%%%%%%%%%%%%

%%%%%%%%%%%%%%%%% APPENDICES %%%%%%%%%%%%%%%%%%%%%

%%%%%%%%%%%%%%%%%%%%%%%%%%%%%%%%%%%%%%%%%%%%%%%%%%
% Don't change these lines if you know whats good for ya.
\bsp	% typesetting comment
\label{lastpage}
\end{document}